\documentclass[conference]{IEEEtran}
\IEEEoverridecommandlockouts

\def\BibTeX{{\rm B\kern-.05em{\sc i\kern-.025em b}\kern-.08em
    T\kern-.1667em\lower.7ex\hbox{E}\kern-.125emX}}

\usepackage{amsmath,amsfonts,amssymb,amsthm}
\usepackage{array}
\usepackage{textcomp}
\usepackage{stfloats}
\usepackage{url}
\usepackage{verbatim}
\usepackage{graphicx}    
\usepackage{epstopdf}    

\usepackage{balance}
\usepackage{cite}
\usepackage[hidelinks]{hyperref}
\usepackage{xcolor}
\usepackage{multirow}
\usepackage{bm}

\usepackage{booktabs}

\usepackage{color}
\usepackage{subcaption}
\usepackage[labelformat=simple]{subcaption}
\usepackage[font=small]{caption}
\usepackage{pdfpages}

\usepackage{tikz}
\usetikzlibrary{tikzmark,decorations.pathreplacing}
\usepackage{rotating}

\usepackage{algorithm}
\usepackage{algorithmic}

\usepackage{fancyhdr}

\usepackage{stfloats}

\usepackage{acronym}

\usepackage{authblk}

\newtheorem*{proof*}{Proof}
\newtheorem*{remark*}{Remark}

\acrodef{ADMM}[ADMM]{alternating direction method of multipliers}

\acrodef{AO}[AO]{alternating optimization}

\acrodef{AoA}[AoA]{angle of arrival}

\acrodef{BS}[BS]{base station}

\acrodef{CRB}[CRB]{Cramér–Rao bound}

\acrodef{ELAA}[ELAA]{extremely large aperture array}

\acrodef{FF}[FF]{far-field}

\acrodef{GHz}[GHz]{gigahertz}

\acrodef{LoS}[LoS]{line-of-sight}

\acrodef{LMMSE}[LMMSE]{linear minimum mean square error}

\acrodef{LS}[LS]{least squares}

\acrodef{MCRB}[MCRB]{misspecified Cramér–Rao bound}

\acrodef{MHz}[MHz]{megahertz}

\acrodef{MLE}[MLE]{maximum likelihood estimation}

\acrodef{mmWave}[mmWave]{millmiter wave}

\acrodef{NF}[NF]{near-field}

\acrodef{NLoS}[NLoS]{non-line-of-sight}

\acrodef{OFDM}[OFDM]{orthogonal frequency division multiplexing}

\acrodef{OMP}[OMP]{orthogonal matching pursuit}

\acrodef{RMSE}[RMSE]{root mean squared error}

\acrodef{SNR}[SNR]{signal-to-noise ratio}

\acrodef{SnS}[SnS]{spatial non-stationary}

\acrodef{TDD}[TDD]{time division duplexing}

\acrodef{THz}[THz]{Terahertz}

\acrodef{UBQP}[UBQP]{unconstrained binary quadratic programming}

\acrodef{UE}[UE]{user equipment}

\acrodef{ULA}[ULA]{uniform linear array}

\acrodef{VR}[VR]{visibility region}

\begin{document}

\title{Joint Near-Field Sensing and Visibility Region Detection with Extremely Large Aperture Arrays \vspace{-3mm}}
\author{Huiping Huang$^{\star}$, Alireza Pourafzal$^{\star}$, Hui Chen$^{\star}$, Musa Furkan Keskin$^{\star}$, Mengting Li$^{\dagger, \star}$, \\ 
Yu Ge$^{\star}$, Fredrik Tufvesson$^{\ddagger}$, Henk Wymeersch$^{\star}$, Xuesong Cai$^{\mathsection,\ddagger}$ \\ \newline \vspace{-3mm} \\
$^{\star}$Chalmers University of Technology ~ $^{\dagger}$Aalborg University ~ $^{\ddagger}$Lund University ~ $^{\mathsection}$Peking University
\vspace{-3mm}
\thanks{This paper is supported by the Vinnova B5GPOS Project under Grant 2022-01640, the HORIZON-MSCA through the project NEAT-6G under Grant 101152670, and the Swedish Research Council (VR) through the project 6G-PERCEF under Grant 2024-04390.}
}

\markboth{}
{Channel estimation and VR detection}

\maketitle

\begin{abstract}
In this paper, we consider near-field localization and sensing with an extremely large aperture array under partial blockage of array antennas, where spherical wavefront and spatial non-stationarity are accounted for. We propose an Ising model to characterize the clustered sparsity feature of the blockage pattern, develop an algorithm based on alternating optimization for joint channel parameter estimation and visibility region detection, and further estimate the locations of the user and environmental scatterers. The simulation results confirm the effectiveness of the proposed algorithm compared to conventional methods. 
\end{abstract}

\section{Introduction}
\label{Introduction}
Sensing, in terms of estimating the state (e.g., position) of \ac{UE} and environmental scatterers, is a key task in wireless networks, which highly overlaps with \textit{channel estimation} in wireless communications \cite{Wymeersch2020, Chen2024August, Gazzah2014}. Sensing (or channel estimation) is mostly based on the planar wavefront assumption in current and past generations of wireless systems \cite{Friedlander2019, Cui2022}. However, it faces serious challenges in future wireless systems, as the array size increases by an order of magnitude and the operating frequencies go to \ac{mmWave} and \ac{THz} bands \cite{Guerra2021, Kosasih2023, Liu2024, Huang2024October}. As a result, the \ac{NF} condition will become increasingly prevalent in various applications. In \ac{NF}, the planar wavefront assumption is invalid, and instead, the spherical wavefront model should be considered. The spherical wavefront model accounts for angle and distance information, which is more complicated than the planar wavefront model. 
Additionally, the entire antenna array is affected by \ac{SnS} due to the \ac{NF} effect and partial blockage \cite{Cai2019, Cai2020, Yuan2023, Li2023} (partial blockage is also referred to as \ac{VR}; see \cite{Han2020, Tian2023, Chen2024, Chen2024July, Zhang2024, Cheng2019August}).

To address the above challenges, various joint channel estimation and \ac{VR} detection methods are proposed in the literature \cite{Han2020, Tian2023, Chen2024, Chen2024July, Zhang2024, Cheng2019August, Iimori2022, Tong2022, Zhu2021, Tang2024, Xu2024}.  These works characterize the \ac{NF} channel response as a product of channel gain, indicator factor, and \ac{NF} steering vector, and can be categorized into deterministic, stochastic, and hybrid channel models.  Deterministic models used in \cite{Han2020, Tian2023, Chen2024, Chen2024July, Zhang2024} are site-specific and they determine the indicator factor as either 0 or 1, while defining the steering vector as a function of angle and distance. Stochastic models are used in \cite{Cheng2019August, Iimori2022, Tong2022} and describe the indicator factor using statistical parameters without explicitly modeling the physical environment. For hybrid models used in \cite{Zhu2021, Tang2024, Xu2024}, the indicator factor is statistically determined as 0 or 1, and the steering vector is a function of angle and distance. The deterministic models and hybrid models are more preferred for positioning and mapping, as the stochastic models do not explicitly capture angle and distance information. Note that all existing studies assume that antenna elements obstructed by blockage (i.e., outside the line-of-sight region) do not contribute to the array's channel response. However, this assumption is practically not true due to the diffraction and penetration effects of obstacles, which have been verified by measurement data as in \cite{Cai2019, Cai2020, Yuan2023, Li2023}.

From the methodology perspective, existing methods can be classified into two groups: non-Bayesian inference \cite{Han2020, Tian2023, Chen2024, Chen2024July, Zhang2024} and Bayesian inference \cite{Cheng2019August, Iimori2022, Tong2022, Zhu2021, Tang2024, Xu2024}. To be specific, non-Bayesian inference methods treat joint channel estimation and \ac{VR} detection as compressive sensing problems, and solve them by orthogonal matching pursuit \cite{Han2020, Chen2024July, Zhang2024}, \ac{MLE} \cite{Tian2023, Chen2024, Chen2024July}, and/or alternating direction method of multipliers \cite{Zhang2024}. These non-Bayesian inference methods \cite{Han2020, Tian2023, Chen2024, Chen2024July, Zhang2024} and one Bayesian inference method \cite{Xu2024} rely on the subarrays, which assume that the subchannel of each subarray can be treated as spatial stationary. However, this assumption may not be practical, especially when there exists partial blockages. For the remaining Bayesian inference methods, \cite{Cheng2019August} and \cite{Tong2022} adopt the Dirichlet process and Bernoulli-Gaussian distribution to model the \ac{VR} indicator factor, respectively, which fail to characterize the clustered sparsity of the \ac{VR} indicator vector; while \cite{Iimori2022}, \cite{Zhu2021}, and \cite{Tang2024} employ nested Bernoulli-Gaussian distribution, three-layer hidden Markov chain, and one-order Markov chain, respectively, to describe the clustered sparsity of the indicator vector. These Markov chains are complicated in terms of modeling the clustered sparsity.

In this paper, we propose a new channel model in partial blockage scenario and an algorithm for joint channel estimation and VR detection. Our main contributions include: (i) We develop a simpler model, i.e., the \textit{Ising} model, to capture the clustered sparsity of the indicator vector. (ii) To account for the data from \cite{Cai2019, Cai2020, Yuan2023, Li2023}, we model the channel response (when it is blocked) as a Gaussian distribution with small variance, which is essentially different from the literature \cite{Han2020, Tian2023, Chen2024, Chen2024July, Zhang2024, Cheng2019August, Iimori2022, Tong2022, Zhu2021, Tang2024, Xu2024}. (iii) We propose an \ac{AO}-based algorithm for joint \ac{NF} sensing and \ac{VR} detection, which performs better than conventional methods, verified by simulations.

\section{Signal Model}
\label{SignalModel}
\begin{figure}[t]
	\vspace*{-2mm}
	\centerline{\includegraphics[width=0.48\textwidth]{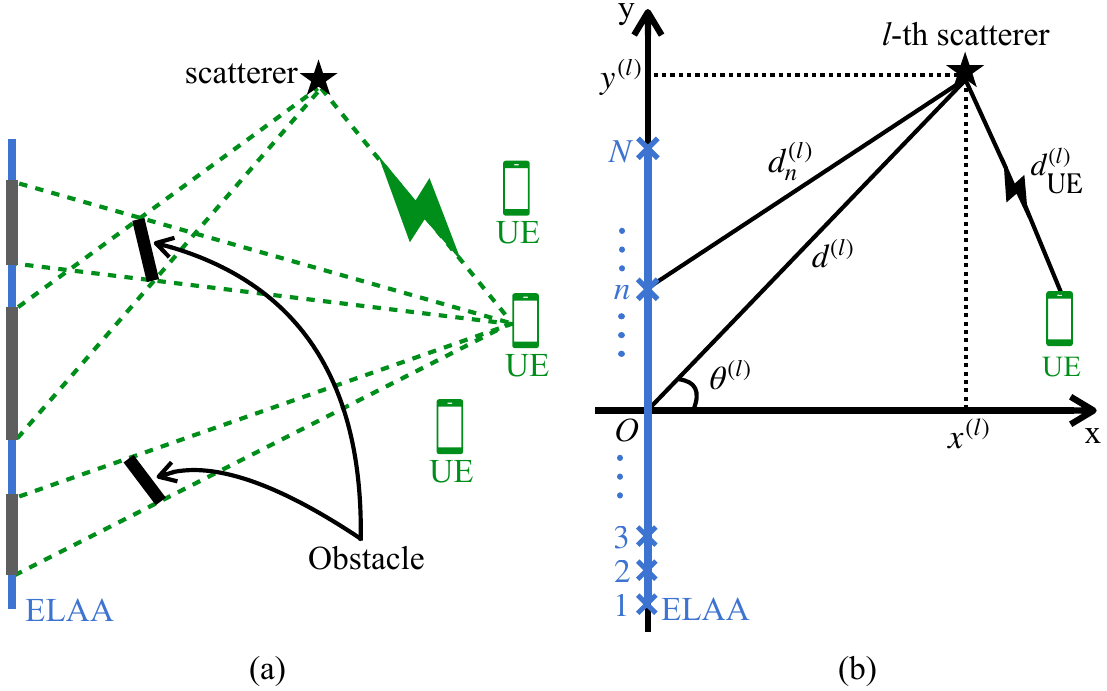}}
	\caption{(a) Illustration of near-field multipath (i.e., multiple scatterers) propagation in the presence of partial blockage. (b) Illustration of position of the $l$-th scatterer in the Cartesian coordinate system.}
	\label{fig_systemmodel}
    \vspace*{-4mm}
\end{figure}

We consider an uplink time division duplexing scenario, as shown in Fig. \ref{fig_systemmodel} (a), where multiple single-antenna \acp{UE} transmit \ac{OFDM} pilot signals to a \ac{BS} equipped with a \ac{ULA} consisting of $N \gg 1$ half-wavelength spaced antennas. Since different \acp{UE} adopt orthogonal pilot sequences during the channel estimation stage, an arbitrary \ac{UE} is considered in the following sections. The pilot signals contain $K$ subcarriers and $T$ symbols. There are obstacles between the \ac{ELAA} and the \ac{UE}, which partially block the \ac{ELAA}. The channel response at the reference point (center point of the \ac{ELAA}) contributed by the $l$-th path reads \cite{Cai2019}
\begin{align}
    x_{k}^{(l)} = g^{(l)}e^{-\jmath 2 \pi f_{k} ({d^{(l)}} + d^{(l)}_{\textrm{UE}})/{c} }, ~~ l = 0, 1, \cdots, L-1,
\end{align}
where $L$ denotes the total number of paths (assumed to be known in this paper), $g^{(l)}$ is the complex channel gain, $f_{k} = f_{c} + k \Delta_{f}$ with $f_{c}$ being the carrier frequency and $\Delta_{f}$ being the subcarrier spacing, $d^{(l)}$ denotes the distance between the $l$-th scatterer and the reference point, $d^{(l)}_{\textrm{UE}}$ denotes the propagation distance from the \ac{UE} to the $l$-th scatterer ($d^{(0)}_{\textrm{UE}} = 0$ for \ac{LoS}), and $c$ is the speed of light. Then, the channel response of the $n$-th antenna at the $k$-th subcarrier and $t$-th snapshot, contributed by the $l$-th path, is given as
\begin{subequations}
\begin{align}
    {x}_{n,k}^{(l)} & = \alpha_{n}^{(l)} x_{k}^{(l)} \frac{d^{(l)}}{d_{n}^{(l)}} e^{\jmath 2 \pi f_{k} {\left(d^{(l)} - d_{n}^{(l)}\right)}/{c} } \\
    & = \alpha_{n}^{(l)} g^{(l)} \frac{d^{(l)}}{d_{n}^{(l)}} e^{- \jmath 2 \pi f_{k} {d_{n}^{(l)}}/{c} } e^{- \jmath 2 \pi f_{k} d^{(l)}_{\textrm{UE}}/c},
\end{align}
\end{subequations}
where $\alpha_{n}^{(l)}$ denotes a stochastic variable characterizing the \ac{SnS} effect, and $d_{n}^{(l)} = \sqrt{ (d^{(l)})^{2} - 2 d^{(l)} \delta_{n} \Delta \sin{(\theta^{(l)})} + \delta_{n}^{2} \Delta^{2} }$ is the distance between the $l$-th scatterer and the $n$-th antenna, with $\theta^{(l)}$ being the \ac{AoA} of the $l$-th path, $\Delta$ being the element-spacing, and $\delta_{n} = \frac{2n - N - 1}{2}$ for $n = 1, 2, \cdots, N$. The relation between the position of the $l$-th scatterer and its channel parameters is shown in Fig. \ref{fig_systemmodel} (b). 

Stacking $x_{n,k}^{(l)}$ for $n = 1, 2, \cdots, N$, into a column vector ${\bf x}_{k}^{(l)} \in \mathbb{C}^{N}$, we obtain ${\bf x}_{k}^{(l)} = g^{(l)} {\bm \alpha}^{(l)} \odot {\bf h}_{k}^{(l)}$, where $\odot$ denotes the Hadamard (entry-wise) product, ${\bm \alpha}^{(l)} \triangleq [\alpha_{1}^{(l)}, \alpha_{2}^{(l)}, \cdots, \alpha_{N}^{(l)}]^{\mathrm{T}} \in \mathbb{C}^{N}$, and ${\bf h}_{k}^{(l)} \in \mathbb{C}^{N}$ is defined as 
\begin{align}
\label{formula_h}
    {\bf h}_{k}^{(l)} = & e^{- \jmath 2 \pi f_{k} d^{(l)}_{\textrm{UE}}/c} \! \left[ \! \frac{d^{(l)}}{d_{1}^{(l)}} e^{- \jmath 2 \pi f_{k} {d_{1}^{(l)}}/{c} }, \cdots \!, \frac{d^{(l)}}{d_{N}^{(l)}} e^{- \jmath 2 \pi f_{k} {d_{N}^{(l)}}/{c} } \! \right]^{ \!\mathrm{T}} \!\!.
\end{align}

The observation data of the \ac{ELAA} at the $k$-th subcarrier and $t$-th snapshot, denoted by ${\bf y}_{k} \in \mathbb{C}^{N}$, can be given as
\begin{align}
\label{obsveration_y}
    {\bf y}_{k,t} = \left( \sum_{l = 0}^{L-1} {\bf x}_{k}^{(l)} \right) s_{k,t} + {\bf n}_{k,t},
\end{align}
where $s_{k,t}$ is the pilot signal (which is equal to 1 without loss of generality), and ${\bf n}_{k,t} \in \mathbb{C}^{N} \sim \mathcal{CN}({\bf 0}, \sigma_{\text{n}}^{2}{\bf I})$ with $\sigma_{\text{n}}^{2}$ denoting the noise variance. Stacking (based on different stacking manners) ${\bf y}_{k,t}$ for all $k$ and $t$, we obtain
\begin{align}
    \label{Simple_sigmal_model_y}
    {\bf{\breve{y}}} = {\bf{\breve{R}}}{\bm \alpha} + {\bf{\breve{n}}}, \quad {\bf{\tilde{y}}} = {\bf{\tilde{R}}}{\bf h} + {\bf{\tilde{n}}}, \quad {\bf{\bar{y}}} = {\bf{\bar{R}}}{\bf g} + {\bf{\bar{n}}},
\end{align}
where the variables are defined in Appendix \ref{Derivative_of_SimpleModel}.

\section{Proposed Model and Method}
\label{ProposedMethod}
\subsection{Proposed Model}
We propose to model the antenna amplitude conditioned on the \ac{VR} as
\begin{align}
\label{prob_alpha_b}
    p({\bm \alpha}^{(l)} | {\bf b}^{(l)}) = \prod_{n = 1}^{N} p(\alpha_{n}^{(l)} | b_{n}^{(l)}),
\end{align}
where ${\bf b}^{(l)} \triangleq [ {b}^{(l)}_{1}, {b}^{(l)}_{2}, \cdots, {b}^{(l)}_{N}]^{\mathrm{T}} \in \mathbb{R}^{N}$, and ${b}_{n}^{(l)}$ indicating if the $n$-th antenna is in the \ac{VR} of the $l$-th path, as  
\begin{align}
\label{eq_bnl}
    b_{n}^{(l)} \!=\! \left\{ \!\!\!
    \begin{array}{l}
    1, \text{~if $n$-th antenna lies in \ac{VR},} \\
    0, \text{~otherwise.}
    \end{array}
    \right.
\end{align}
In \eqref{prob_alpha_b}, the amplitude at each antenna is modeled as
\begin{align}
\label{prob_alpha_n}
    p(\alpha_{n}^{(l)} | b_{n}^{(l)}) = (1 - b_{n}^{(l)}) \mathcal{CN}(0, \sigma_{\text{b}}^{2}) + b_{n}^{(l)} \delta(\alpha_{n} - 1),
\end{align}
where $\sigma_{\text{b}}^{2}$ denotes the amplitude variance in the blockage region, and $\delta(\cdot)$ is the Dirac delta function. When the $n$-th antenna lies in the \ac{VR}, then $b_{n}^{(l)} = 1$ and its amplitude follows $\delta(\alpha_{n} - 1)$, meaning that $\alpha_{n} = 1$ with probability one; otherwise, $b_{n}^{(l)} = 0$ and its amplitude follows $\mathcal{CN}(0, \sigma_{\text{b}}^{2})$. Note that when $b_{n}^{(l)} = 1$ for all $n$ and $l$, the observation reduces to ${\bf x}^{(l)}_{k} = g^{(l)}{\bf h}^{(l)}_{k}$, which corresponds to the general case of deterministic channel model without considering the \ac{SnS}. 

To proceed, we approximate the Dirac delta function by a complex Gaussian distribution with a very small variance $\sigma_{\text{v}}^{2}$, and rewrite \eqref{prob_alpha_n} as
\begin{align}
\label{prob_alpha_n_app}
    p(\alpha_{n}^{(l)} | b_{n}^{(l)}) & \approx (1 - b_{n}^{(l)}) \mathcal{CN}(0, \sigma_{\text{b}}^{2}) + b_{n}^{(l)} \mathcal{CN}(1, \sigma_{\text{v}}^{2}) \nonumber \\
    & = \mathcal{CN}(b_{n}^{(l)}, (1-b_{n}^{(l)})\sigma_{\text{b}}^{2} + b_{n}^{(l)}\sigma_{\text{v}}^{2}).
\end{align}
Since whether the antenna lies in the \ac{VR} is related to its nearby antennas, we propose to utilize an \textit{Ising model} (a standard type of Markov random fields \cite{Bishop2006}) to characterize the clustered sparsity of ${\bf b}^{(l)}$, which is given as  
\begin{align}
\label{prob_alpha}
    p({\bf b}^{(l)}) = \frac{1}{A} e^{ - \left( \displaystyle \sum_{(n,m) \in \mathcal{E}} \beta_{nm}^{(l)} b_{n}^{'(l)} b_{m}^{'(l)} + \sum_{n = 1}^{N} \gamma_{n}^{(l)} b_{n}^{'(l)} \right) },
\end{align}
where $b^{'(l)}_{n} \triangleq 2b_{n}^{(l)} - 1$ are substitution variables, $\mathcal{E}$ is the set of edges representing neighboring antenna pairs, $\beta_{nm}^{(l)}$ denotes the interaction strength between the $n$-th and $m$-th antennas, $\gamma_{n}^{(l)}$ is the individual strength parameter for the $n$-th antenna, and $A$ is the normalization parameter. A detailed explanation of the proposed Ising model is given in Appendix \ref{IsingModel}. 

\subsection{Proposed Method}
The joint distribution based on \eqref{obsveration_y} can be given by
\begingroup
\allowdisplaybreaks
\begin{align}
\label{likelihoofFun}
    & p({\bf y}, \! \{\!{\bm \alpha}^{\!(l)}\!\} \!, \{\!{\bf b}^{\!(l)}\!\}\!, \{\!{\bf g}^{\!(l)}\!\}; \{\!\theta^{\!(l)}\!\}\!, \{\!d^{\!(l)}\!\}\!, \{\!d^{\!(l)\!}_{\textrm{UE}}\!\}\!) \nonumber \\
    = & p({\bf y} | \{\! {\bm \alpha}^{\!(l)} \!\}\!, \{\!{\bf b}^{\!(l)}\!\}\!, \{\!{\bf g}^{\!(l)}\!\}; \{\!\theta^{\!(l)}\!\}\!, \{\!d^{\!(l)}\!\}\!, \{\!d^{\!(l)\!}_{\textrm{UE}}\!\}\!) \! \prod_{l = 0}^{L\!-\!1} \! p(\!{\bm \alpha}^{\!(l)} | {\bf b}^{\!(l)}\!) p(\!{\bf b}^{\!(l)}\!),
\end{align}
\endgroup
where $p({\bf y} | \{{\bm \alpha}^{(l)}\}, \{{\bf b}^{(l)}\}, \{{\bf g}^{(l)}\}; \{\theta^{(l)}\}, \{d^{(l)}\}\!, \{d^{(l)}_{\textrm{UE}}\}) = \mathcal{CN}(\sum_{l = 0}^{L-1}{\bf x}^{(l)}, \sigma_{\text{n}}^{2}{\bf I})$. Substituting the results in \eqref{prob_alpha_b}, \eqref{prob_alpha_n_app} and \eqref{prob_alpha} into \eqref{likelihoofFun} yields the negative log-likelihood function, $- \log ~\! p({\bf y}, \{{\bm \alpha}^{(l)}\}, \{{\bf b}^{(l)}, \{{\bf g}^{(l)}\}; \{\theta^{(l)}\}, \{d^{(l)}\}, \{d^{(l)}_{\textrm{UE}}\}\}) = f_{1} + f_{2} + f_{3} + L \log A$, where 
\begin{subequations}
\label{f_functions}
\begin{align}
    \label{f_funtion_1}
    f_{1} & \triangleq \frac{1}{\sigma_{\text{n}}^{2}} \left\| {{\bf y}} - {{\bf R}}{\bf w}\right\|_{2}^{2}, \\
    f_{2} & \triangleq \sum_{l = 0}^{L-1} \sum_{n = 1}^{N} \frac{ (\alpha_{n}^{(l)} - b_{n}^{(l)})^{2} }{ (1 - b_{n}^{(l)})\sigma_{\text{b}}^{2} + b_{n}^{(l)}\sigma_{\text{v}}^{2} }, \\
    \label{f_function_3}
    f_{3} & \triangleq \! \sum_{l = 0}^{L-1} \! \left( \sum_{(n,m) \in \mathcal{E}} \! \beta_{nm}^{(l)} b_{n}^{'(l)} b_{m}^{'(l)} + \sum_{n = 1}^{N} \gamma_{n}^{(l)} b_{n}^{'(l)} \! \right),
\end{align}
\end{subequations}
and $L \log A$ is unrelated to the unknown channel parameters $\{\theta^{(l)}, d^{(l)}, d^{(l)}_{\textrm{UE}}, g^{(l)}, {\bm \alpha}^{(l)}\}$ and the VR indicator vectors $\{{\bf b}^{(l)}\}$. In addition, ${{\bf y}}$, ${\bf R}$, and ${\bf w}$ in \eqref{f_funtion_1} equal to ${\bf{\breve{y}}}$, ${\bf{\breve{R}}}$, and ${\bm \alpha}$, respectively, when the first model in \eqref{Simple_sigmal_model_y} is considered; ${{\bf y}}$, ${\bf R}$, and ${\bf w}$ equal to ${\bf{\tilde{y}}}$, ${\bf{\tilde{R}}}$, and ${\bf h}$, respectively, when the second model in \eqref{Simple_sigmal_model_y} is considered; and ${{\bf y}}$, ${\bf R}$, and ${\bf w}$ equal to ${\bf{\bar{y}}}$, ${\bf{\bar{R}}}$, and ${\bf g}$, respectively, when the third model in \eqref{Simple_sigmal_model_y} is considered. Consequently, the estimates of channel parameters and \ac{VR} indicator vectors can be obtained via
\begin{align}
\label{problem_channel}
    \left\{ \!\! \{ \! {\widehat \theta}^{(l)} \!\} \!, \{\!{\widehat d}^{(l)} \!\} \!, \{\!{\widehat d}^{(l)}_{\textrm{UE}} \!\}, \{\!{\widehat{g}}^{(l)} \!\} \!, \{\!{\widehat{\bm \alpha}}^{(l)}\!\} \!, \{\!{\widehat{\bf b}}^{(l)}\!\} \!\! \right\} = {\arg\min} ~ f_{1} \!+\! f_{2} \!+\! f_{3}.
\end{align}

The hyperparameters of the Ising model can be estimated by measurement data, and the estimations of channel parameters, VR indicator vector, and positions of all scatterers are presented as follows.

\subsubsection{Estimation of Channel Parameters and \ac{VR} Indicator}
By introducing the following auxiliary variables defined as: 
\begingroup
\allowdisplaybreaks
\begin{align*}
    {\bf b} & \! \triangleq {\bf 1}_{T} \otimes \left[ ({\bf b}^{(0)})^{\mathrm{T}}, ({\bf b}^{(1)})^{\mathrm{T}}, \cdots, ({\bf b}^{(L-1)})^{\mathrm{T}} \right]^{\mathrm{T}} \mathbb{R}^{NLT \times 1} \\
    {\bf b}^{'} & \! \triangleq 2{\bf b} - {\bf 1}_{NLT} \in \{-1, 1\}^{NLT \times 1}, \\
    {\bf q} & \! \triangleq \sigma_{\text{b}}^{2} {\bf 1}_{NLT} + \left(\sigma_{\text{v}}^{2} - \sigma_{\text{b}}^{2}\right){\bf b} \in \mathbb{R}^{NLT \times 1}, \\
    {\bf Q} & \! \triangleq \mathrm{diag}( {\bf q} ) \in \mathbb{R}^{NLT \times NLT}, \\
    {\widetilde{\bf E}}^{(l)} & \! \triangleq \begin{bmatrix}
        ~ 0      & \beta_{12}^{(l)}     & \cdots & \beta_{1N}^{(l)} \\
        ~ \beta_{21}^{(l)}      & 0     & \cdots & \beta_{2N}^{(l)} \\
        ~ \vdots & \vdots  & \ddots      & \vdots \\
        ~ \beta_{N1}^{(l)}      & \beta_{N2}^{(l)}  & \cdots  & 0 \\
        \end{bmatrix} \! \in \! \mathbb{R}^{N \times N}, \\
    {\bf E} & \! \triangleq {\bf I}_{T} \otimes \mathrm{blkdiag} \! \left( \! {\widetilde{\bf E}}^{(0)} \!, {\widetilde{\bf E}}^{(1)} \!, \cdots \!, {\widetilde{\bf E}}^{(L-1)} \! \right) \! \in \! \mathbb{R}^{NLT \times NLT}, \\
    {\widetilde{\bm \gamma}}^{(l)} & \! \triangleq \! \left[ \gamma_{1}^{(l)} , \gamma_{2}^{(l)} , \cdots, \gamma_{N}^{(l)} \right]^{\mathrm{T}} \in \mathbb{R}^{N \times 1}, \\
    {\bm \gamma} & \! \triangleq \! {\bf 1}_{T} \otimes \! \left[ \! \left( \! {\widetilde{\bm \gamma}}^{(0)} \! \right)^{\!\mathrm{T}} \!, \left( \! {\widetilde{\bm \gamma}}^{(1)} \! \right)^{\!\mathrm{T}} \!, \cdots \!, \left( \! {\widetilde{\bm \gamma}}^{(L-1)} \! \right)^{\!\mathrm{T}}\right]^{\mathrm{T}} \!\! \in \! \mathbb{R}^{NLT \times 1},
\end{align*}
\endgroup
we reformulate: $f_{1} = \frac{1}{\sigma_{\text{n}}^{2}} \| {{\bf y}} - {{\bf R}}{\bf w}\|_{2}^{2}$, $f_{2} = \| {\bf Q}^{-\frac{1}{2}} ({\bm \alpha} - {\bf b} )\|_{2}^{2}$, and $f_{3} = \frac{1}{2}{\bf b}^{'\mathrm{T}}{\bf E}{\bf b}^{'} + {\bm \gamma}^{\mathrm{T}}{\bf b}^{'}$.
Problem \eqref{problem_channel} is solved by the following two-step strategy:
\begin{align}
\label{problem_Handalpha}
    \text{Step 1:} ~ \{\widehat{\bf g}, \widehat{\bf h}, \widehat{\bm \alpha}, \widehat{\bf b}\} = \! \underset{ \{{\bf g}, {\bf h}, {\bm \alpha}, {\bf b}\}}{\arg\min} ~\! f_{1} \!+\! f_{2} \!+\! f_{3} ~~ \mathrm{s.t.} ~ {\bf b} \! \in \! \{0 , 1\}^{NLT},
\end{align}
\begin{align}
\label{problem_channelandalpha}
    \text{Step 2:} ~ & \left\{ \! \{ \! {\widehat \theta}^{(l)} \! \}, \{ \! {\widehat d}^{(l)} \!\}, \{\! {\widehat d}^{(l)}_{\textrm{UE}} \! \} \! \right\} \gets \widehat{\bf h}, ~~ {\widehat{\bm \alpha}}^{(l)}_{t} = \left[\widehat{\bm \alpha}\right]_{t , (l-1)N+1 : lN}, \nonumber \\
    & {\widehat{g}}^{(l)} = \left[ {\widehat{\bf g}} \right]_{l}, ~~ {\widehat{\bf b}}^{(l)} = \left[\widehat{\bf b}\right]_{(l-1)N+1 : lN}.
\end{align}

\begin{algorithm}[b]
\caption{Proposed \ac{AO}-Based Algorithm for Solving \eqref{problem_Handalpha}}
\label{Alg_AO}
\begin{algorithmic}[1]
\STATE \textbf{Input:} ${\bf y}$ (${\bf{\bar y}}$, ${\bf{\tilde y}}$, and ${\bf{\breve y}}$), $\mathbf{E}$, $\bm{\gamma}$, $\sigma_{\text{n}}^{2}$, $\sigma_{\text{b}}^{2}$, $\sigma_{\text{v}}^{2}$, $I$, $\epsilon$
\STATE \textbf{Output:} ${\bf g}_{(i+1)}$, ${\bf h}_{(i+1)}$, ${\bm{\alpha}}_{(i+1)}$, ${\bf b}_{(i+1)}$

\STATE Initialize: ${\bf g}_{(0)}$, ${\bf h}_{(0)}$, ${\bm \alpha}_{(0)}$, ${\bf b}_{(0)}$

\FOR{ $i = 0, 1, \cdots, I$ }
    \STATE \!\!${\bf{\bar R}} \gets \eqref{R_bar}$, ${\bf g}_{(i+1)} = ({\bf{\bar R}}^{\mathrm{H}}{\bf{\bar R}})^{-1}{\bf{\bar R}}^{\mathrm{H}}{\bf{\bar y}}$
    \STATE \!\!${\bf{\tilde{R}}} \gets \eqref{R_tilde}$, ${\bf h}_{(i+1)} = ({\bf{\tilde R}}^{\mathrm{H}}{\bf{\tilde R}})^{-1}{\bf{\tilde R}}^{\mathrm{H}}{\bf{\tilde y}}$
    \STATE \!\!${\bf{\breve R}} \!\!\gets\!\! \eqref{R_breve}$, ${\bm \alpha}_{(\!i+1\!)} \!\!=\! {\bm \mu}_{\alpha} \!+\! {\bm \Sigma}_{\alpha} \! {\breve{\bf R}}^{\!\mathrm{H}}\!(\!{\breve{\bf R}}{\bm \Sigma}_{\alpha}\!{\breve{\bf R}}^{\!\mathrm{H}} \!\!+\! \sigma_{\text{n}}^{2}{\bf I}\!)^{\!-\!1}\!(\!{\breve{\bf y}} \!-\! {\breve{\bf R}}\!{\bm \mu}_{\alpha}\!)$
    \STATE \!\!${\bf b}_{(i+1)} \gets \eqref{problem_b_reformation}$
    \IF{ $\frac{ |\mathrm{Obj}_{(i+1)} - \mathrm{Obj}_{(i)}| }{ | \mathrm{Obj}_{(i)} | } \leq \epsilon$ }
        \STATE \textbf{break}
    \ENDIF
\ENDFOR
\end{algorithmic}
\end{algorithm}

\subsubsection*{\underline{Solve \eqref{problem_Handalpha}}}
\label{SubsubSection_Solve_Handalpha}
Parameters ${\bf g}$, ${\bf h}$, ${\bm \alpha}$, and ${\bf b}$ can be estimated under the \ac{AO} framework. 

\begin{itemize}
    \item Estimate ${\bf g}$: Given ${\bf{\bar{y}}}$ and ${\bf{\bar{R}}}$ defined in \eqref{y_bar} and \eqref{R_bar}, respectively, the \ac{LS} solution for $\min_{{\bf g}} ~ \| {\bf{\bar{y}}} - {\bf{\bar{R}}}{\bf g} \|_{2}^{2}$ can be expressed as $\widehat{\bf g} = ({\bf{\bar R}}^{\mathrm{H}}{\bf{\bar R}})^{-1}{\bf{\bar R}}^{\mathrm{H}}{\bf{\bar y}}$.
    \item Estimate ${\bf h}$: Given ${\bf{\tilde{y}}}$ and ${\bf{\tilde{R}}}$ defined in \eqref{y_tilde} and \eqref{R_tilde}, respectively, the \ac{LS} solution for $\min_{{\bf h}} ~ \| {\bf{\tilde y}} - {\bf{\tilde R}}{\bf h} \|_{2}^{2}$ can be expressed as $\widehat{\bf h} = ({\bf{\tilde R}}^{\mathrm{H}}{\bf{\tilde R}})^{-1}{\bf{\tilde R}}^{\mathrm{H}}{\bf{\tilde y}}$.
    \item Estimate ${\bm \alpha}$: To solve $\min_{{\bm \alpha}} ~ \| {\bf{\breve y}} - {\bf{\breve R}}{\bm \alpha} \|_{2}^{2}$ with ${\bf{\breve{y}}}$ and ${\bf{\breve{R}}}$ defined in \eqref{y_breve} and \eqref{R_breve}, respectively, the \ac{LMMSE} estimator is adopted. The solution can be expressed as $\widehat{\bm \alpha} = {\bm \mu}_{\alpha} + {\bm \Sigma}_{\alpha}{\breve{\bf R}}^{\mathrm{H}}({\breve{\bf R}}{\bm \Sigma}_{\alpha}{\breve{\bf R}}^{\mathrm{H}} + \sigma_{\text{n}}^{2}{\bf I})^{-1}({\breve{\bf y}} - {\breve{\bf R}}{\bm \mu}_{\alpha})$, where ${\bm \mu}_{\alpha} = {\bf b}$ and ${\bm \Sigma}_{\alpha} = \mathrm{diag}(\sigma_{\text{b}}^{2}({\bf 1} - {\bf b}) + \sigma_{\text{v}}^{2}{\bf b})$ according to \eqref{prob_alpha_n_app}.
    \item Estimate ${\bf b}$: The subproblem 
    $\min_{{\bf b} \in \{0, 1\}^{NLT}} ~ ({\bm \alpha} \!-\! {\bf b})^{\mathrm{H}}{\bf Q}^{-1}({\bm \alpha} \!-\! {\bf b}) + \frac{1}{2}{\bf b}^{'\mathrm{T}}{\mathbf E}{\mathbf b}^{'} + {\bm \gamma}^{\mathrm{T}}{\mathbf b}^{'}$ can be relaxed to an unconstrained binary quadratic programming as
\begin{align}
\label{problem_b_reformation}
    \min_{{\bf b}} ~\!\! 2{\bf b}^{\mathrm{T}}\!{{\bf E}}{\bf b} \!+\! {\bf r}^{\mathrm{T}}{\bf b}, ~ \mathrm{s.t.} \! \left\{ \!\!\!\!
    \begin{array}{l}
        {\bf 0} \preceq {\bf b} \preceq {\bf 1}_{NLT}, \\
        \left({\bf I}_{NLT} \!-\! \mathrm{diag}({\bf b})\!\right) \! {\bf b} \! \preceq \! {\bm \eta},
    \end{array}
    \right.
\end{align}
where ${\bm \eta} \in \mathbb{R}^{NLT}$ is predefined and ${\bf r} = {\bf r}_{1} + {\bf r}_{2} + {\bf r}_{3}$ with ${\bf r}_{1} \triangleq \left[ \frac{|\alpha_{1} - 1|^{2}}{\sigma_{\text{v}}^{2}} \!-\! \frac{|\alpha_{1}|^{2}}{\sigma_{\text{b}}^{2}}, \cdots, \frac{|\alpha_{NLT} - 1|^{2}}{\sigma_{\text{v}}^{2}} \!-\! \frac{|\alpha_{NLT}|^{2}}{\sigma_{\text{b}}^{2}} \right]^{\mathrm{T}}$, ${\bf r}_{2} \triangleq -2{\bf E}^{\mathrm{T}}{\bf 1}$, and ${\bf r}_{3} \triangleq 2{\bm \gamma}$. This problem can be solved via numerical optimization solvers.
\end{itemize}

\subsubsection*{\underline{Solve \eqref{problem_channelandalpha}}}
\label{SubsubSection_Solve_channelandalpha}
Channel gain, antenna amplitude, and \ac{VR} for each path and snapshot can be estimated by \eqref{problem_channelandalpha}. The distances and \ac{AoA} of each path can be estimated based on \ac{MLE} using $\widehat{\bf h}$ obtained from \eqref{problem_Handalpha}. To be specific, we extract ${\widehat{\bf h}}_{k}^{(l)} \in \mathbb{C}^{N}$ for all $K$ subcarriers from $\widehat{\bf h}$. Then, we have
\begin{align}
\label{estimate_theta_d_ML}
    \left( {\widehat{d}}^{(l)}_{\textrm{UE}} , {\widehat{d}}^{(l)} , {\widehat{\theta}}^{(l)} \right) = \arg\min_{d_{\textrm{UE}}, d, {\theta}} ~ \sum_{ k =1 }^{K} \left\| \widehat{\bf h}^{(l)}_{k} - {\bf h}_{k}(d_{\textrm{UE}}, d, {\theta}) \right\|_{2}^{2},
\end{align}
where ${\bf h}_{k}(d_{\textrm{UE}}, d, {\theta})$ is a function of $d_{\textrm{UE}}$, $d$, and $\theta$ as in \eqref{formula_h}.

\subsubsection{Estimation of Positions of \ac{UE} and Scatterers}
After we obtain the estimates of distance and \ac{AoA} of each path, the position of the $l$-th scatterer, i.e., $(\widehat{x}^{(l)} , \widehat{y}^{(l)})$, can be calculated as: $\widehat{x}^{(l)} = \widehat{d}^{(l)}\cos{(\widehat{\theta}^{(l)})}, ~ \widehat{y}^{(l)} = \widehat{d}^{(l)}\sin{(\widehat{\theta}^{(l)})}$.

\begin{figure}[t]
	\vspace*{-2mm}
	\centerline{\includegraphics[width=0.5\textwidth]{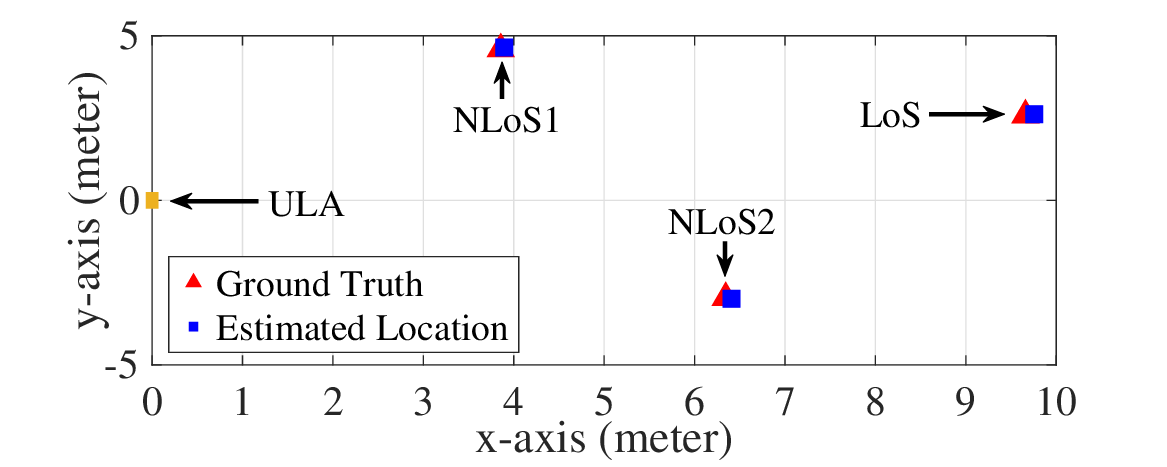}}
	\caption{Estimate of locations of $L = 3$ scatterers.}
	\label{Loc_scatterer}
    \vspace*{-2mm}
\end{figure}

\section{Simulations}
\label{simulation}
We consider a \ac{ULA} of $N = 100$ antennas. A \ac{UE} transmits \ac{OFDM} signals, with carrier frequency being $f_{c} = 30$ GHz (i.e., wavelength $\lambda = 0.01$ m), $K = 4$ subcarriers, $T = 4$ snapshots, and bandwidth $2.88$ MHz. The simulation parameters are compliant with 3GPP \cite{3GPP}. The \ac{ELAA} element-spacing is set to be half of the wavelength, i.e., $\Delta = \frac{\lambda}{2} = 0.005$ m. The Fraunhofer distance is $d_{F} \triangleq \frac{2D^{2}}{\lambda} \approx 50$ m. We consider $L = 3$ paths, including one \ac{LoS} and two \ac{NLoS} paths. Other simulation parameters are given below.
\begin{itemize}
    \item \ac{LoS}: \ac{UE} at $(d^{(0)}, {\theta}^{(0)}) = (10 \text{~m}, 15^{\circ})$, and the antennas from index $\{75\}$ to index $\{80\}$ are blocked.
    \item \ac{NLoS}1: The scatterer at $(d^{(1)}, {\theta}^{(1)}) = (6 \text{~m} , 50^{\circ})$, and the antennas from index $\{11\}$ index $\{14\}$ are blocked.
    \item \ac{NLoS}2: The scatterer at $(d^{(2)}, {\theta}^{(2)}) = (7 \text{~m} , -25^{\circ})$, and the antennas from index $\{34\}$ to index $\{38\}$ are blocked.
\end{itemize}

\begin{figure}[t]
	\vspace*{-2mm}
	\centerline{\includegraphics[width=0.45\textwidth]{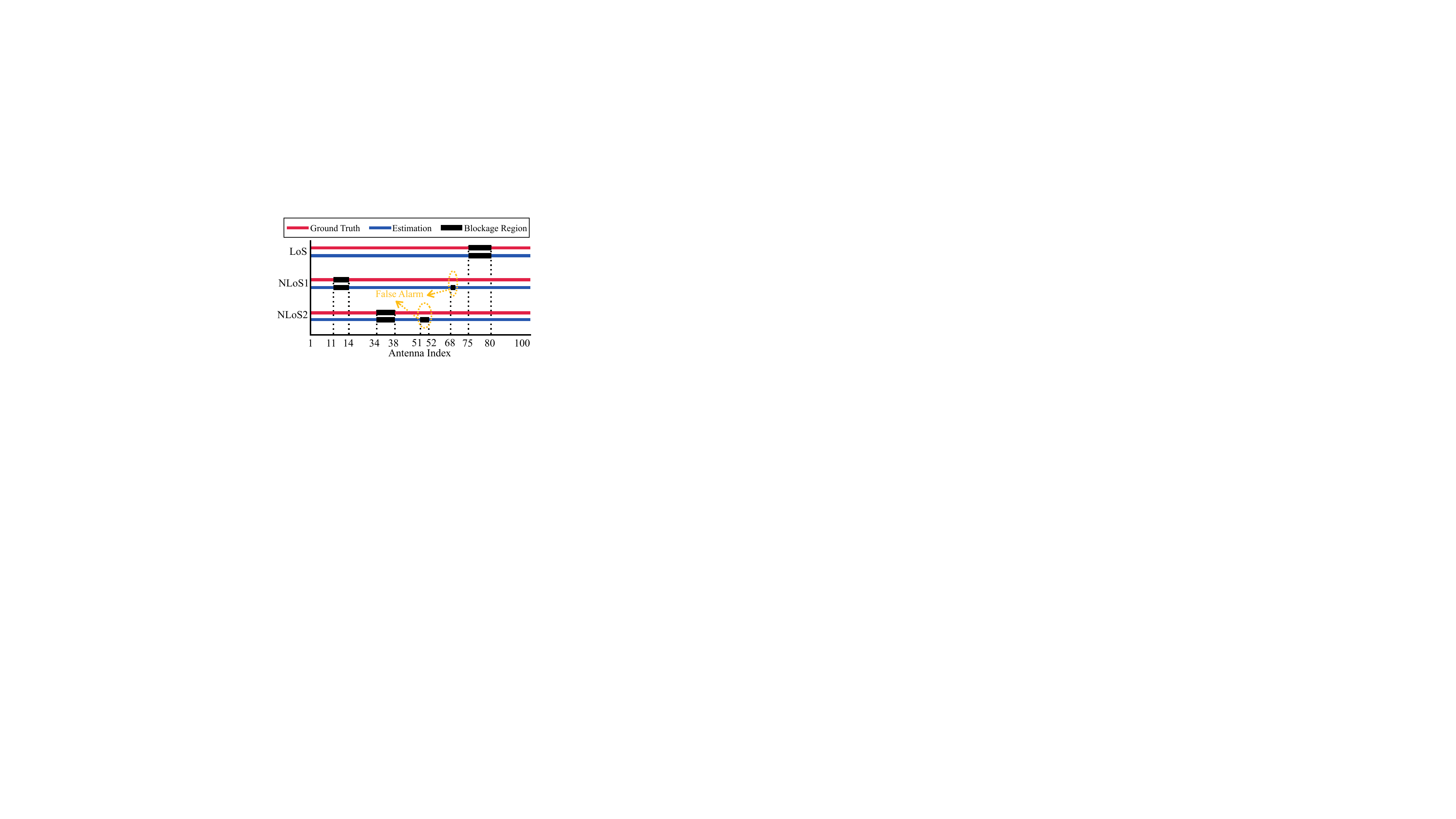}}
	\caption{Blockage region detection with SNR = 10 dB.}
	\label{fig_VR_detection}
    \vspace*{0mm}
\end{figure}

\begin{figure}[t]
	\vspace*{-2mm}
	\centerline{\includegraphics[width=0.5\textwidth]{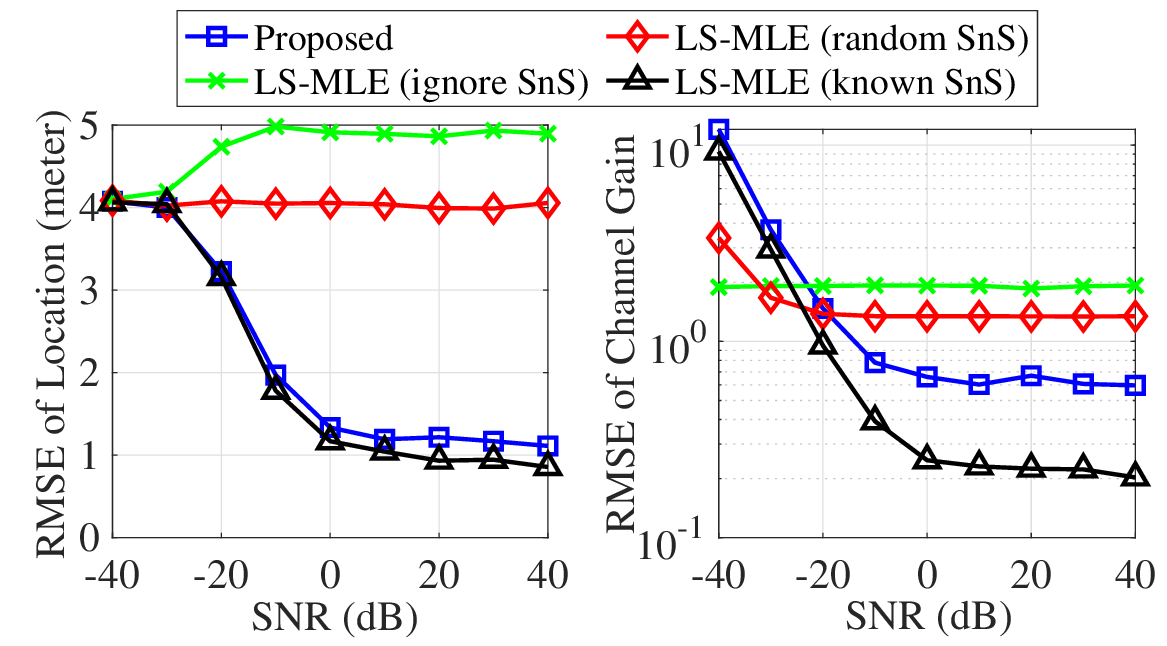}}
	\caption{\ac{RMSE} of location (left) and channel gain (right) versus \ac{SNR}.}
	\label{RMSE_versus_SNR}
    \vspace*{-2mm}
\end{figure}

The estimated locations and ground truth of $L = 3$ scatterers are depicted in Fig. \ref{Loc_scatterer}, which shows that the estimated locations are very close to the ground truth. The estimated \ac{VR} are displayed in Fig. \ref{fig_VR_detection}, showing that the proposed algorithm can detect all blocked antennas, and has false alarm for the two \ac{NLoS} paths. The \ac{RMSE} of location and channel gain versus \ac{SNR} are plotted in Figs. \ref{RMSE_versus_SNR} (left) and \ref{RMSE_versus_SNR} (right), respectively. It is seen that the proposed method outperforms two \ac{LS}-\ac{MLE}-based methods, without considering the \ac{SnS} and with random \ac{SnS}, respectively. The \ac{LS}-\ac{MLE}-based method with known \ac{SnS} is used as a benchmark, and the proposed method is slightly worse than it.

\section{Conclusion}
\label{conclusion}
We investigated near-field localization and sensing with an extremely large aperture array (ELAA). We proposed an Ising model to characterize the clustered sparsity of the blockage pattern of ELAA and developed an alternating optimization-based algorithm for joint channel parameter estimation and visibility region detection. The simulation results indicated that the proposed algorithm achieves better performance than traditional methods.

\appendices
\section{Derivation of Equation \eqref{Simple_sigmal_model_y}}
\label{Derivative_of_SimpleModel}
We denote the realization of the SnS effect ${\alpha}_{n}^{(l)}$ at the $t$-th snapshot as ${\alpha}_{n,t}^{(l)}$, and define ${\bm \alpha}_{t}^{(l)} \triangleq [\alpha_{1,t}^{(l)}, \alpha_{2,t}^{(l)}, \cdots, \alpha_{N,t}^{(l)}]^{\mathrm{T}} \in \mathbb{C}^{N}$. Inserting ${\bf x}_{k,t}^{(l)} = g^{(l)} {\bm \alpha}_{t}^{(l)} \odot {\bf h}_{k}^{(l)}$ into \eqref{obsveration_y} yields: ${\bf y}_{k,t} = \left( {\bf g}^{\mathrm{T}} \! \otimes \! {\bf I}_{N} \right) \! \mathrm{diag}({\bm \alpha}_{t}) {\bf h}_{k} + {\bf n}_{k,t}$, where $\otimes$ is the Kronecker product, $\mathrm{diag}(\cdot)$ generates a diagonal matrix with the argument as its main diagonal, and
\begin{align*}
    {\bf g} & \triangleq [g^{(0)}, g^{(1)}, \cdots, g^{(L-1)}]^{\mathrm{T}} \in \mathbb{C}^{L}, \\
    {\bm \alpha}_{t} & \triangleq [ ({\bm \alpha}_{t}^{(0)})^{\mathrm{T}}, ({\bm \alpha}_{t}^{(1)})^{\mathrm{T}}, \cdots, ({\bm \alpha}_{t}^{(L-1)})^{\mathrm{T}} ]^{\mathrm{T}} \in \mathbb{C}^{NL}, \\
    {\bf h}_{k} & \triangleq [ ({\bf h}_{k}^{(0)})^{\mathrm{T}}, ({\bf h}_{k}^{(1)})^{\mathrm{T}}, \cdots, ({\bf h}_{k}^{(L-1)})^{\mathrm{T}} ]^{\mathrm{T}} \in \mathbb{C}^{NL}.
\end{align*}
Stack ${\bf y}_{k,t}$ for all $k$ into a matrix, as
\begin{align*}
    {\bf Y}_{t} = \left( {\bf g}^{\mathrm{T}} \otimes {\bf I}_{N} \right) \mathrm{diag}({\bm \alpha}_{t})[{\bf h}_{1}, {\bf h}_{2}, \cdots, {\bf h}_{K}] + {\bf N}_{t},
\end{align*}
where ${\bf N}_{t} \triangleq [{\bf n}_{1,t}, {\bf n}_{2,t}, \cdots, {\bf n}_{K,t}] \in \mathbb{C}^{N \times K}$. Vectorize ${\bf Y}_{t}$ as
\begin{align*}
    {\bf y}_{t} \triangleq \mathrm{vec}({\bf Y}_{t}) = \left( [{\bf h}_{1}, {\bf h}_{2}, \cdots, {\bf h}_{K}]^{\mathrm{T}} \circledast \left( {\bf g}^{\mathrm{T}} \otimes {\bf I}_{N} \right) \right) {\bm \alpha}_{t} + {\bf n}_{t},
\end{align*}
where $\circledast$ denotes the Khatri-Rao product and ${\bf n}_{t} \triangleq \mathrm{vec}({\bf N}_{t}) \in \mathbb{C}^{NK}$. Stack ${\bf y}_{t}$ for all $t$ as a vector ${\bf{\breve{y}}} \in \mathbb{C}^{NKT}$, as
\begin{align}
\label{y_breve}
    {\bf{\breve{y}}} \triangleq [{\bf y}_{1}^{\mathrm{T}}, {\bf y}_{2}^{\mathrm{T}}, \cdots, {\bf y}_{T}^{\mathrm{T}}]^{\mathrm{T}} = {\bf{\breve{R}}}{\bm \alpha} + {\bf{\breve{n}}},
\end{align}
where
\begin{subequations}
\begin{align}
    \label{R_breve}
    {\bf{\breve{R}}} & \triangleq {\bf I}_{T} \!\otimes\! \left( [{\bf h}_{1}, \cdots \!, {\bf h}_{K}]^{\mathrm{T}} \! \circledast \! \left( {\bf g}^{\mathrm{T}} \! \otimes \! {\bf I}_{N} \!\right) \!\right) \!\in\! \mathbb{C}^{NKT \times NLT}, \\
    {\bm \alpha} & \triangleq [ {\bm \alpha}_{1}^{\mathrm{T}}, {\bm \alpha}_{2}^{\mathrm{T}}, \cdots, {\bm \alpha}_{T}^{\mathrm{T}} ]^{\mathrm{T}} \in \mathbb{C}^{NLT},  \\
    \label{n_breve}
    {\bf{\breve{n}}} & \triangleq [ {\bf n}_{1}^{\mathrm{T}}, {\bf n}_{2}^{\mathrm{T}}, \cdots, {\bf n}_{T}^{\mathrm{T}} ]^{\mathrm{T}} \in \mathbb{C}^{NKT}.
\end{align}
\end{subequations}

Inserting ${\bf x}_{k,t}^{(l)} = g^{(l)} {\bm \alpha}_{t}^{(l)} \odot {\bf h}_{k}^{(l)}$ into \eqref{obsveration_y} yield: ${\bf y}_{k,t} = \left( {\bf g}^{\mathrm{T}} \! \otimes \! {\bf I}_{N} \right) \! \mathrm{diag}({\bf h}_{k}) {\bm \alpha}_{t} + {\bf n}_{k,t}$. Stack ${\bf y}_{k,t}$ for all $t$ into a matrix:
\begin{align*}
    {\bf Y}_{k} = \left( {\bf g}^{\mathrm{T}} \otimes {\bf I}_{N} \right) \mathrm{diag}({\bf h}_{k}) [{\bm \alpha}_{1}, {\bm \alpha}_{2}, \cdots, {\bm \alpha}_{T}] + {\bf N}_{k},
\end{align*}
where ${\bf N}_{k} \triangleq [{\bf n}_{k,1}, {\bf n}_{k,2}, \cdots, {\bf n}_{k,T}] \in \mathbb{C}^{N \! \times \! T}$. Vectorize ${\bf Y}_{k}$ as ${\bf y}_{k} \triangleq \mathrm{vec}({\bf Y}_{k}) = \left( [{\bm \alpha}_{1}, {\bm \alpha}_{2}, \cdots, {\bm \alpha}_{T}]^{\mathrm{T}} \! \circledast \! \left( {\bf g}^{\mathrm{T}} \! \otimes \! {\bf I}_{N} \right) \right) {\bf h}_{k} + {\bf n}_{k}$, with ${\bf n}_{k} \! \triangleq \! \mathrm{vec}({\bf N}_{k}) \! \in \! \mathbb{C}^{NT}$. Stack ${\bf y}_{k}$ for all $k$ as ${\bf{\tilde{y}}} \in \mathbb{C}^{NTK}$:
\begin{align}
    \label{y_tilde}
    {\bf{\tilde{y}}} \triangleq [{\bf y}_{1}^{\mathrm{T}}, {\bf y}_{2}^{\mathrm{T}}, \cdots, {\bf y}_{K}^{\mathrm{T}}]^{\mathrm{T}} = {\bf{\tilde{R}}}{\bf h} + {\bf{\tilde{n}}},
\end{align}
where 
\begin{subequations}
\begin{align}
    \label{R_tilde}
    {\bf{\tilde{R}}} & \triangleq {\bf I}_{K} \! \otimes \! \left( [{\bm \alpha}_{1}, \cdots\!, {\bm \alpha}_{T}]^{\mathrm{T}} \!\! \circledast \! \left( {\bf g}^{\mathrm{T}} \!\! \otimes \! {\bf I}_{N} \! \right) \! \right) \!\in\! \mathbb{C}^{NTK \times NLK}, \\
    \label{h}
    {\bf h} & \triangleq [{\bf h}_{1}^{\mathrm{T}}, {\bf h}_{2}^{\mathrm{T}}, \cdots, {\bf h}_{K}^{\mathrm{T}}]^{\mathrm{T}} \in \mathbb{C}^{NLK}, \\
    \label{n_tilde}
    {\bf{\tilde{n}}} & \triangleq [{\bf n}_{1}^{\mathrm{T}}, {\bf n}_{2}^{\mathrm{T}}, \cdots, {\bf n}_{K}^{\mathrm{T}}]^{\mathrm{T}} \in \mathbb{C}^{NTK}.
\end{align}
\end{subequations}

Inserting ${\bf x}_{k,t}^{(l)} = g^{(l)} {\bm \alpha}_{t}^{(l)} \odot {\bf h}_{k}^{(l)}$ into \eqref{obsveration_y} also yields: ${\bf y}_{k,t} = {\bf R}_{k,t}{\bf g} + {\bf n}_{k,t}$, where ${\bf R}_{k,t} \triangleq [{\bm \alpha}_{t}^{(0)}\!, {\bm \alpha}_{t}^{(1)}\!, \! \cdots \!, {\bm \alpha}_{t}^{(L-1)}] \odot [{\bf h}_{k}^{(0)}\!, {\bf h}_{k}^{(1)}\!, \! \cdots \!, {\bf h}_{k}^{(L-1)}]$. Stack ${\bf y}_{k,t}$ for all $k$ and $t$ as a vector ${\bf{\bar{y}}} \in \mathbb{C}^{NKT}$, as
\begin{align}
    \label{y_bar}
    {\bf{\bar{y}}} \triangleq [ {\bf y}_{1,1}^{\mathrm{T}}, \cdots \!, {\bf y}_{1,T}^{\mathrm{T}}, \cdots \!, {\bf y}_{K,1}^{\mathrm{T}}, \cdots \!, {\bf y}_{K,T}^{\mathrm{T}} ]^{\mathrm{T}} \!\!= {\bf{\bar{R}}}{\bf g} + {\bf{\bar{n}}},
\end{align}
\begin{subequations}
where
\begin{align}
    \label{R_bar}
    {\bf{\bar{R}}} & \triangleq [ {\bf R}_{1,1}^{\mathrm{T}}, \! \cdots \!, {\bf R}_{1,T}^{\mathrm{T}}, \! \cdots \!, {\bf R}_{K,1}^{\mathrm{T}}, \! \cdots \!, {\bf R}_{K,T}^{\mathrm{T}} ]^{\mathrm{T}} \!\! \in \! \mathbb{C}^{N\!K\!T \!\times\! L} \!, \\
    \label{g}
    {\bf g} & \triangleq [g^{(0)}, g^{(1)}, \cdots, g^{(L-1)}]^{\mathrm{T}} \in \mathbb{C}^{L}, \\
    \label{n_bar}
    {\bf{\bar{n}}} & \triangleq [ {\bf n}_{1,1}^{\mathrm{T}}, \cdots \!, {\bf n}_{1,T}^{\mathrm{T}}, \cdots \!, {\bf n}_{K,1}^{\mathrm{T}}, \cdots \!, {\bf n}_{K,T}^{\mathrm{T}} ]^{\mathrm{T}} \in \mathbb{C}^{NKT}.
\end{align}
\end{subequations}

\section{Explanation of the Ising Model}
\label{IsingModel}
First of all, the substitution variables $b_{n}^{'}$ are equal to either $-1$ or $1$, since $b_{n} = 0$ or $1$. We analyze the key components in the Ising model, i.e., $\beta_{nm}b_{n}^{'}b_{m}^{'} + \gamma_{n}b_{n}^{'} + \gamma_{m}b_{m}^{'}$, where we ignore the path index. Our goal is to minimize $\beta_{nm}b_{n}^{'}b_{m}^{'} + \gamma_{n}b_{n}^{'} + \gamma_{m}b_{m}^{'}$. Clearly, a negative $\beta_{nm}$ will enforce $b_{n}^{'} = b_{m}^{'} = 1$ or $b_{n}^{'} = b_{m}^{'} = -1$; and a positive $\beta_{nm}$ will enforce ($b_{n}^{'} = 1$ and $b_{m}^{'} = -1$) or ($b_{n}^{'} = -1$ and $b_{m}^{'} = 1$). On the other hand, a negative $\gamma_{n}$ (reps. $\gamma_{m}$) will enforce $b_{n}^{'} = 1$ (resp. $b_{m}^{'} = 1$); while a positive $\gamma_{n}$ (resp. $\gamma_{m}$) will enforce $b_{n}^{'} = -1$ (resp. $b_{m}^{'} = -1$). This indicates that the proposed Ising model can lead to clustered sparsity of ${\bf b}$ due to the terms $\beta_{nm}b_{n}^{'}b_{m}^{'}$.

\balance
\bibliographystyle{Sub/myIEEEtran}      
\bibliography{Sub/refs}

\end{document}